\documentclass[a4paper,doublecol]{epl2} 

\usepackage{amsmath}
\usepackage{amssymb}

\usepackage{color}

\title{Signatures of protein structure in the cooperative gating \\ of mechanosensitive ion channels}
\shorttitle{Signatures of protein structure in cooperative gating} 

\author{Osman Kahraman,\inst{1} William S. Klug\inst{2} \and Christoph A. Haselwandter\inst{1}}
\shortauthor{Osman Kahraman \etal}

\institute{                    
\inst{1}Departments of Physics \& Astronomy and Biological Sciences, University of Southern California \\ Los Angeles, CA 90089, USA\\
\inst{2}Department of Mechanical and Aerospace Engineering, University of California, Los Angeles \\
Los Angeles, CA 90095, USA}
\abstract{Membrane proteins deform the surrounding lipid bilayer, which can
lead to membrane-mediated interactions between neighboring proteins. 
Using the mechanosensitive channel of large conductance (MscL) as a model system, we demonstrate how the observed differences in protein structure can affect membrane-mediated interactions and cooperativity among membrane proteins. We find that distinct oligomeric states of MscL lead to distinct gateway states for the clustering of MscL, and predict signatures of MscL structure and spatial organization in the cooperative gating of MscL.
Our modeling approach establishes a quantitative relation between the observed shapes and cooperative function of membrane~proteins.
}

\begin{document}

\maketitle
\section{Introduction}
The biological function of membrane proteins is often influenced by an interplay
between protein structure \cite{bowie2005solving} and the mechanical properties of the surrounding lipid bilayer \cite{milescu2009interactions,brohawn2012crystal,schmidt2012mechanistic,anishkin2013stiffened}. One way to quantify the regulation of protein function by lipid bilayer mechanics is to consider the hydrophobic interface between the membrane protein and the lipid bilayer. A hydrophobic mismatch at the bilayer-protein interface can induce thickness deformations of the bilayer membrane \cite{Huang1986,AndersenKoeppe,jensen04} which typically extend over distances comparable to the protein size \cite{Phillips2009}. The energetics of thickness deformations can be captured by a simple continuum elastic model \cite{Huang1986, AndersenKoeppe,jensen04,Phillips2009,lundbaek06}, in which the lipid bilayer is represented by a thin elastic body and proteins are represented as rigid inclusions. In the crowded membrane environment
most relevant for living cells \cite{dupuy08,takamori06,linden12}, deformation fields of neighboring membrane proteins are expected to overlap, leading to membrane-mediated interactions between proteins \cite{Phillips2009,harroun99}. 

\begin{figure}
\onefigure[width=7cm]{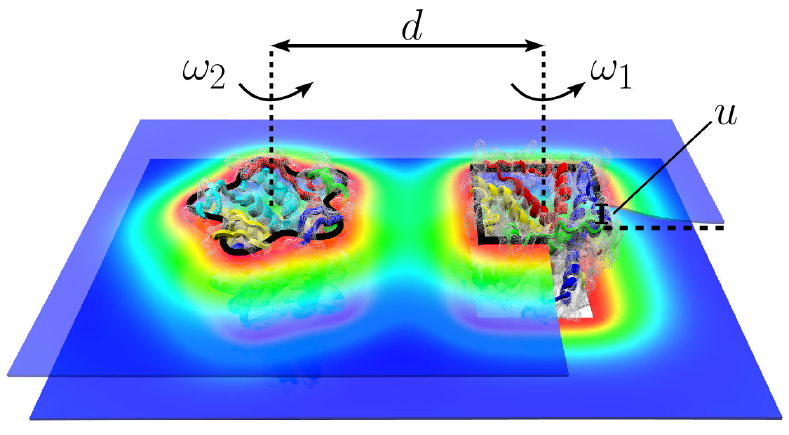} 
\caption{\label{fig:pentamer} (Color online)
Thickness deformations induced by a pair of pentameric \cite{Chang1998}
(left protein structure)
and tetrameric \cite{Liu2009} (right protein structure) MscL in the face-on configuration.
The deformation profile depends on MscL structure, center-to-center separation, $d$, and orientation, $\omega_{1,2}$. 
The pentameric and tetrameric MscL structures correspond to Protein Data Bank accession numbers 2OAR and 3HZQ, respectively.
The diameter of MscL in the closed (shown here) and open states is $\sim\!\!5$ nm and $\sim\!\!7$ nm, respectively \cite{Ursell2008}. 
 \label{fig:definitions}}
\end{figure}

Mechanosensitive ion channels provide an experimental model system
\cite{kung10,booth07,perozo06,sachs10} for studying the connection between
membrane protein conformation and the mechanical properties of lipid bilayers. In particular, experiments on the bacterial mechanosensitive channel of large conductance (MscL) \cite{Perozo2002a,Perozo2006,Booth2007,Hamill2001,Markin2004,Chiang2004,Belyy2010,Anishkin2005}
have helped to establish a quantitative relation \cite{Wiggins2004,Wiggins2005,Ursell2008,CAH2013b} between the MscL gating probability, 
the applied membrane tension, and lipid bilayer material properties such as bilayer hydrophobic thickness.
Moreover, membrane-mediated interactions between MscL \cite{Ursell2007,CAH2013a} have been observed to yield MscL clusters \cite{grage2011bilayer} and cooperative effects \cite{grage2011bilayer,Nomura2012}.
Despite recent breakthroughs in structural membrane biology \cite{White2009}, the physiologically relevant oligomeric state of MscL is still a source of
debate \cite{Haswell2011,Dorwart2010,Iscla2011,Gandhi2010}. Early work suggested a hexameric symmetry \cite{Saint1998}, but high-resolution protein crystallography has pointed to pentameric \cite{Chang1998} (fig.~\ref{fig:pentamer} left panel) as well as tetrameric \cite{Liu2009} (fig.~\ref{fig:pentamer} right panel) MscL stoichiometries. Do the observed membrane-mediated interactions between MscL bear signatures of MscL symmetry? Do distinct MscL stoichiometries induce distinct cooperative gating characteristics? In this letter we introduce a finite element modeling approach that allows a systematic survey of membrane-mediated interactions for the various structures and symmetries proposed for MscL. 
We explore quantitatively how differences in MscL structure and symmetry are reflected in the anisotropy of membrane-mediated interactions (fig.~\ref{fig:pentamer}), yielding substantial shifts in the cooperative MscL gating probability with the stoichiometry and spatial configuration of MscL.

\section{Modeling MscL function}
\begin{figure}[t]
\onefigure[width=7.5cm]{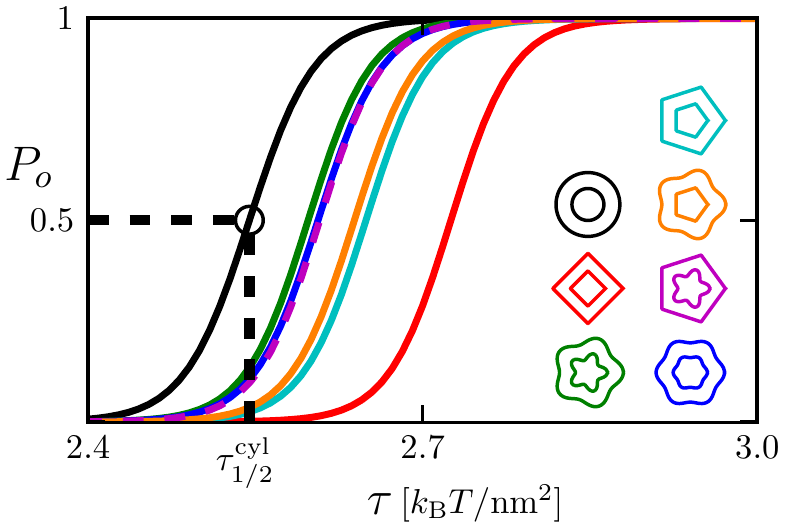}
\caption{ (Color online)
Gating curves for the closed and open MscL shapes suggested by structural studies \cite{Liu2009,Chang1998,Saint1998,Sukharev2001a,Sukharev2001b} and the cylinder
model of MscL \cite{Wiggins2004,Wiggins2005,Ursell2008,Ursell2007} shown in the insets (closed states superimposed on the interior of open states). The gating tension $\tau_{1/2}$ corresponds to a probability $P_o=1/2$ of being in the open state, and is indicated by
$\tau_{1/2}^\text{cyl}$ for the cylinder model of MscL. Estimated gating tensions (curves from left to right):  $\tau_{1/2}=2.545$
$k_\text{B}T/$nm$^2$, $2.597$ $k_\text{B}T/$nm$^2$, $2.606$ $k_\text{B}T/$nm$^2$, $2.608$ $k_\text{B}T/$nm$^2$, $2.639$ $k_\text{B}T/$nm$^2$, $2.649$ $k_\text{B}T/$nm$^2$,
and $2.725$ $k_\text{B}T/$nm$^2$.
The purple line is dashed for ease of visualization.
\label{fig:gating_one}} 
\end{figure}

Although the gating of MscL with increasing membrane tension \cite{Perozo2002a,Perozo2006,Booth2007,Hamill2001,Markin2004,Chiang2004,Belyy2010,Anishkin2005}
is a complex process involving many intermediate states, the competition between open and closed states of MscL can be captured by a simple two-state Boltzmann model \cite{Chiang2004,Hamill2001, Markin2004,Belyy2010,Perozo2002a,Anishkin2005}.
The central quantity in this model is the channel opening probability
\begin{align}
 P_o = \frac{1}{1+e^{\beta(\Delta G -\tau \Delta A)} },
 \label{eq:proba}
\end{align}
where $\Delta G$ and $\Delta A$ are the free energy and area differences between open and closed states of MscL, $\tau$ denotes the membrane tension, and $\beta=1/k_\text{B} T$, in which $k_\text{B}$ is Boltzmann's constant and $T$ is the temperature.  Equation~(\ref{eq:proba}) implies that MscL
can exist in closed or open conformations, with the competition between these
two states governed by membrane tension (fig.~\ref{fig:gating_one}). The transition energy $\Delta G$ generally depends \cite{Wiggins2004,Wiggins2005} on the internal protein free energy and on the membrane deformation energy of MscL. 
We focus here on the membrane contributions to $\Delta G$, which allows us to dissect the effect of protein shape on membrane-mediated interactions and cooperative gating of MscL. Indeed, the basic phenomenology of MscL gating can already be understood
\cite{Wiggins2004,Wiggins2005,Phillips2009} by considering the membrane contributions to $\Delta G$, which can take a similar magnitude as experimental estimates of the total MscL transition energy \cite{Perozo2002a,Markin2004,Chiang2004,Belyy2010}.

We quantify $\Delta G$ using the standard elastic theory of membranes
\cite{Huang1986, AndersenKoeppe,jensen04,Phillips2009,lundbaek06},
in which the lipid bilayer is represented by two fields  $h_+(x,y)$ and $h_-(x,y)$ defining the upper and lower boundaries of the hydrophobic bilayer core at
the Cartesian coordinates $(x,y)$. For MscL, the dominant membrane deformation is generally due to thickness mismatch \cite{Phillips2009,Wiggins2005,Wiggins2004}, which is governed by the elastic
energy
\begin{equation} \label{eq:elastic_energy}
{ G=\frac{1}{2}}\int \text{d}x \text{d}y {\textstyle\left\{K_b (\nabla^2 u)^2+K_t \left(\frac{u}{a}\right)^2+\tau
\left[2 \frac{u}{a}+(\nabla u)^2 \right] \right\}}\,,
\end{equation}
where the thickness deformation field is given by ${\textstyle u(x,y) = \frac{1}{2}\left[ h_+(x,y) - h_-(x,y) - 2a \right]}$,
in which $2a$ denotes the thickness of the unperturbed lipid bilayer, $K_b$ denotes the bending rigidity of the lipid bilayer, and $K_t$ is the stiffness associated with thickness deformations. 
The term $2 \tau u/a$ in eq.~(\ref{eq:elastic_energy})  accounts for stretching deformations tangent to the monolayer surfaces (\textit{i.e.}, changes to the areal density of lipids) \cite{Ursell2007,Ursell2008} while the term $\tau (\nabla u)^2$ accounts for changes in the projection of the membrane area onto the plane due to the sloping of the monolayers \cite{Wiggins2004,Wiggins2005}.
We use the typical bilayer parameter values \cite{Ursell2008} $K_b = 20$ $k_\text{B} T$, $K_t = 60$ $k_\text{B} T/$nm$^2$, and $a=1.75$~nm.
For the boundary shapes of MscL we follow ref. \cite{CAH2013b} and consider coarse-grained representations 
obtained from fits to the proposed hydrophobic cross sections of MscL \cite{Liu2009,Chang1998,Sukharev2001a,Sukharev2001b,Saint1998}.

\begin{figure*}[t]\center
 \includegraphics[width=6.cm]{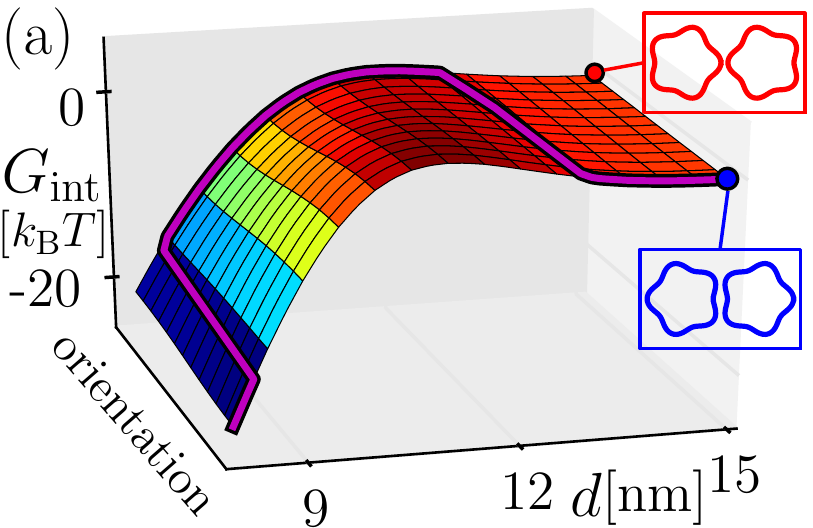} \hspace{1.2cm}
 \includegraphics[width=6.cm]{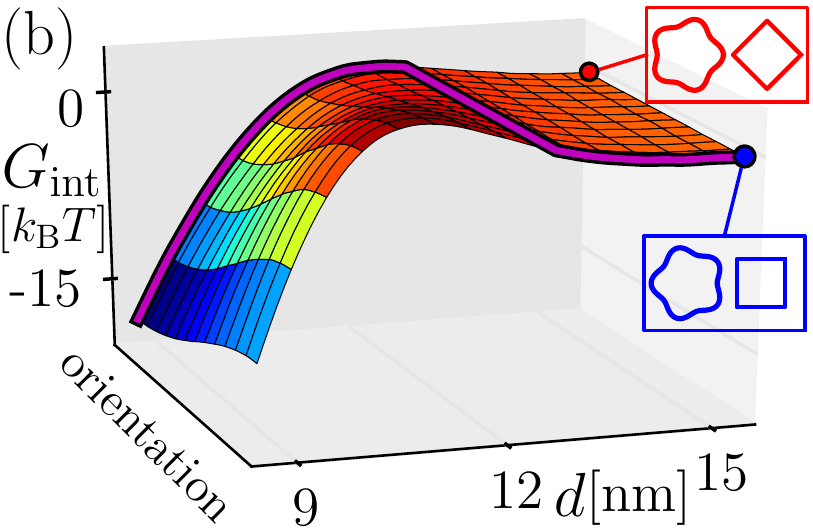} \\ \vspace{0.5cm}
 \includegraphics[width=6.cm]{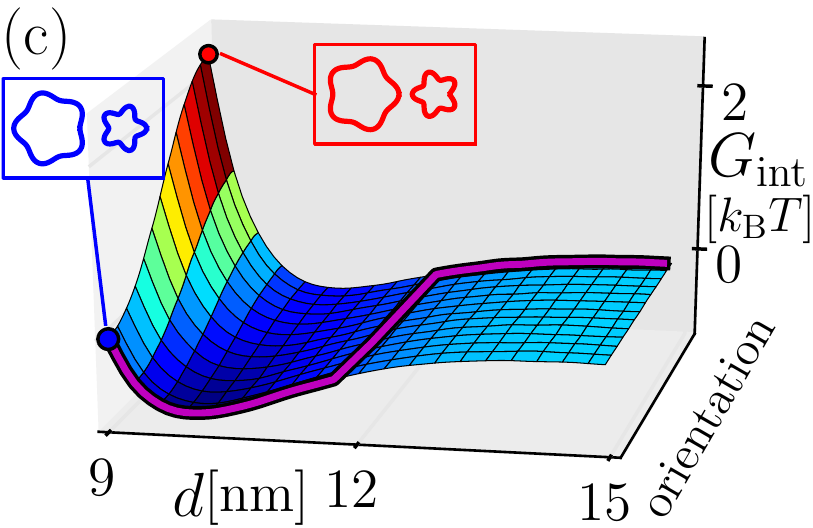} \hspace{1.2cm}
 \includegraphics[width=6.cm]{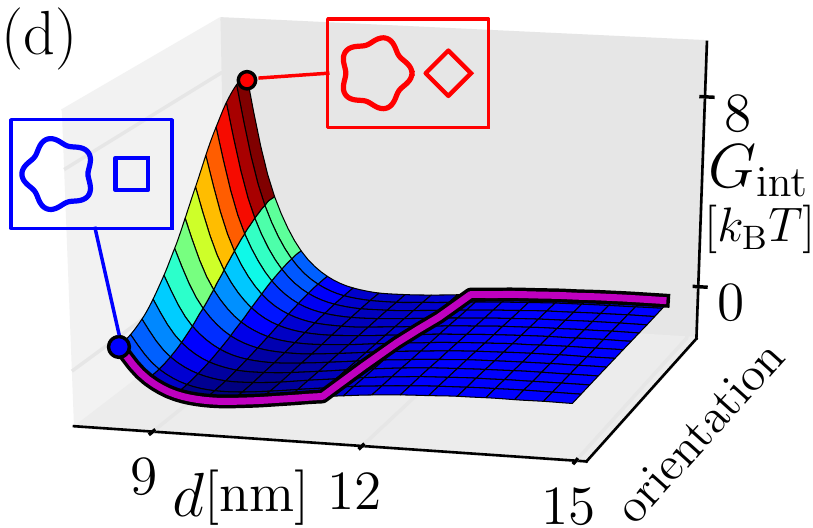}
\caption{(Color online)
Interaction energy, $G_\textrm{int}$, between pentameric and tetrameric MscL as a function of protein separation and orientation, for which we consider
symmetric rotations from the face-on (blue insets) to the tip-on (red insets) configuration. Purple curves show the minimum energy configuration for each $d$. (a) Two pentameric MscL in the open state \cite{Sukharev2001a}. (b) Open pentameric \cite{Sukharev2001a} and open tetrameric \cite{Liu2009} MscL.
(c) Open \cite{Sukharev2001a} and closed \cite{Chang1998} pentameric~MscL.
(d) Open pentameric \cite{Sukharev2001a} and closed tetrameric \cite{Liu2009} MscL.
\label{fig:interactions}}
\end{figure*}

\section{Calculation of thickness deformations}
The thickness deformation footprint of MscL can be quantified by minimization of eq.~(\ref{eq:elastic_energy}) subject to the boundary conditions at the bilayer-protein interface. For cylindrical, non-interacting membrane proteins this calculation is readily performed analytically 
\cite{Huang1986,AndersenKoeppe,jensen04,Phillips2009,lundbaek06}. For non-cylindrical
or interacting membrane proteins, analytic series solutions \cite{CAH2013a,CAH2013b} and numerical finite difference schemes \cite{harroun99b,Ursell2007,Mondal2012} have been developed. 
It has been found \cite{CAH2013a} that, at the small protein separations most relevant for crowded cell membranes, finite difference approaches
show a discrepancy with exact analytic solutions, presumably due to the very small grid sizes required to capture strong angular variations.
Moreover, analytic solutions are difficult to construct for complicated protein shapes such as provided by MscL.
To overcome these challenges we developed a novel finite element approach that is able to capture the thickness
deformations induced by the complicated protein shapes observed in structural studies. Finite element schemes are highly versatile and provide good numerical accuracy, especially for integration domains with complicated
boundaries \cite{bathe2006finite}, which makes them ideally suited for analyzing bilayer-protein interactions for complicated protein structures.
The combined presence of both first and second derivatives in the energy in eq. \ref{eq:elastic_energy} places special demands on the finite element formulation. While standard Lagrange interpolation is adequate to compute the thickness stretch and gradient terms, it fails to produce conforming curvatures at element interfaces. For the bending terms we therefore use the discrete Kirchhoff triangle formulation---an efficient and accurate method for computing curvatures \cite{Batoz1980,Bathe1981}.

\section{Gating of non-interacting MscL}
In the dilute limit, \textit{i.e.}, for isolated MscL, the primary effect of anisotropic channel shape on gating is to shift the gating tension $\tau_{1/2}$ corresponding to  $P_o(\tau_{1/2}) = 1/2$  (figure~\ref{fig:gating_one}).
For the cylinder model of MscL (black curve) we find a gating tension $\tau_{1/2}^\text{cyl}\approx2.5 \,k_\text{B}T/$nm$^2$. For the typical bilayer properties \cite{Ursell2008} used here, the proposed MscL structures \cite{Liu2009,Chang1998,Sukharev2001a,Sukharev2001b,Saint1998}
imply higher gating tensions than the cylinder model of MscL. Indeed, the
tetrameric model of MscL (red curve in fig.~\ref{fig:gating_one})
yields the highest gating tension among all the proposed structural models, while the lowest gating tension is obtained for the pentameric clover-leaf model of MscL (green curve in fig.~\ref{fig:gating_one}). The hexameric (blue curve in fig.~\ref{fig:gating_one}) and pentameric polygonal (cyan
curve in fig.~\ref{fig:gating_one}) models gate at intermediate tensions, with the pentameric polygonal model gating at a slightly higher tension than the hexameric model. We also consider the case of pentameric MscL with a polygonal shape
in the closed state and a clover-leaf shape in the open state (orange curve
in fig.~\ref{fig:gating_one}) as well as the reverse case of a closed clover-leaf
shape and and an open polygonal shape (purple curve in fig.~\ref{fig:gating_one}).
We predict that the gating tensions associated with these models are very
close to the gating tensions of the pentameric polygonal and hexameric
MscL models, respectively. To a good approximation, all the gating curves
in fig.~\ref{fig:gating_one} are parallel to each other, implying that the sensitivity of MscL, defined as the derivative of the open probability with respect to membrane tension \cite{Ursell2007}, does not vary considerably among the proposed MscL models, with the largest deviation being $\sim4\%$.
The qualitative trends in the relative gating tensions of tetrameric, pentameric,
and hexameric MscL in fig.~\ref{fig:gating_one} generally agree with the corresponding perturbative analytic results obtained at first order \cite{CAH2013b}, but fig.~\ref{fig:gating_one} implies values of $\tau_{1/2}$ which are lower by $\sim0.1$ to $\sim0.4$ $k_\text{B}T/$nm$^2$.

\section{Gateways to MscL dimerization}
The thickness deformation fields induced by neighboring MscL overlap, yielding
membrane-mediated interactions which extend over several nanometers (fig.~\ref{fig:pentamer}). Figure~\ref{fig:interactions}
shows the MscL interaction energy as a function of pair configurations. We explore the configuration space for the center-to-center distance $d$ to a minimum edge-to-edge separation $\sim\!\!0.8$~nm 
set by steric constraints on lipid size, and for symmetric rotations $\omega_1=\pi/5-\omega_2$. 
It may be conjectured that, at small protein separations, the minimal edge-to-edge distance may provide a more direct parameterization of the interaction energy than the center-to-center distance. However, we found that plotting interaction energies in terms of edge separations yields no additional insight, and even makes the interpretation of results less straightforward.
Starting with the face-on configuration, intermediate orientations are obtained through a counter-clockwise rotation of the MscL models on the right and a simultaneous clockwise rotation of the MscL models on the left. Thus, the energy profiles shown in fig.~\ref{fig:interactions} correspond to a projection of the overall energy landscape of MscL interactions to symmetric rotations.
Based on the complete interaction potentials of MscL, which allow for arbitrary
protein orientations, we indeed find that the symmetric rotations considered
in fig.~\ref{fig:interactions} provide a good approximation of the minimum
energy configurations. Alternative projections, which are energetically slightly less favorable, would correspond to simultaneous clockwise or counter-clockwise rotations of both channels.

Figure~\ref{fig:interactions}(a)
shows the interaction potential for the pentameric clover-leaf model of MscL
in the open state. We find that, at MscL separations
greater than $d\approx 13$ nm, membrane-mediated interactions are negligible. For intermediate values of $d$, from $d \approx 10$~nm to $d \approx 13$~nm, membrane-mediated interactions are weakly repulsive, with an interaction strength of the order of $1 \, k_\text{B} T$. For separations smaller than $d \approx 10$ nm, membrane-mediated interactions are strongly attractive, yielding dimerization of identical MscL. The minimum-energy orientation of
pentameric MscL changes abruptly as a function of $d$ (see purple curve in fig.~\ref{fig:interactions}(a)), yielding a sequence
of characteristic ``gateway states'' for dimerization of MscL. For separations
greater than $d\approx 13$~nm, the face-on configuration is slightly favorable over the tip-on
configuration. 
However, as membrane-mediated interactions become repulsive at $d \approx 13$~nm, the tip-on orientation
becomes most favorable. Finally, at separations smaller than $d\approx9$~nm, the face-on configuration minimizes the elastic energy of thickness deformations. These conclusions are consistent with results obtained using a perturbative analytic approach \cite{CAH2013a}. We find the same sequence of gateway states for polygonal models of pentameric MscL as well as for hexameric MscL, provided both channels are either in the closed or the open state. Furthermore, the sequence of gateway states in fig.~\ref{fig:interactions}(a) is also
obtained for mixed pairs of pentameric and hexameric MscL. 

We predict a sequence of gateway states distinct from fig.~\ref{fig:interactions}(a) for MscL pairs involving tetrameric MscL, such as tetrameric and pentameric MscL in the open state (fig.~\ref{fig:interactions}(b)), or two tetrameric MscL or tetrameric and hexameric MscL. In particular, the transition at small $d$ from the tip-on to the face-on configuration does not occur for MscL pairs involving tetrameric MscL, which we attribute to the
small internal angle associated with tetrameric vertices. For tetrameric MscL such a transition can only be precipitated
by steric constraints, which become effective for protein separations smaller than those considered in fig.~\ref{fig:interactions}. Moreover, we find that the interaction strength is different for different oligomeric states of MscL, with stronger interactions for tetrameric MscL, by $\sim\!\!3 \,k_\text{B}T$ for the open state, and weaker interactions for hexameric MscL, by $\sim\!\!\!2 \,k_\text{B}T$ for the open state, compared to the pentameric clover-leaf model of MscL.

A distinctive sequence of gateway states is also obtained if one channel is in the open state and the other channel is in the closed state (fig.~\ref{fig:interactions}(c,d)), with the face-on configuration being most favorable at separations smaller than $d\approx12$~nm and the tip-on configuration being most favorable at separations greater than $d\approx12$~nm. In this case we predict the same sequence of gateway states for systems
composed of identical (fig.~\ref{fig:interactions}(c)) and distinct (fig.~\ref{fig:interactions}(d)) oligomeric states. Thus, variation of the MscL oligomeric state modifies the sequence of gateway states if both channels are open or closed, but not if one channel is open and the other channel is closed. Consistent with previous studies \cite{Ursell2007,grage2011bilayer,CAH2013a} we find strong repulsion
between open and closed MscL at small separations and weak attraction at intermediate separations. These results can be understood by noting that the closed and open states of MscL have distinct hydrophobic thicknesses, yielding frustration of membrane deformations at small separations but, due to the overshoot of thickness deformations away
from the bilayer-protein boundary \cite{Huang1986,dan1993membrane}, an energetically favorable overlap of membrane deformations of the same sign at intermediate separations. In contrast, dimerization of MscL of the same hydrophobic thickness reduces the overall membrane deformation footprint of the two channels and, hence, is energetically favorable \cite{Ursell2007,grage2011bilayer,CAH2013a}, while the overshoot in thickness deformations
\cite{Huang1986,dan1993membrane} yields repulsion between identical MscL at intermediate separations.

\begin{figure}[t]
 \onefigure[width=7.5cm]{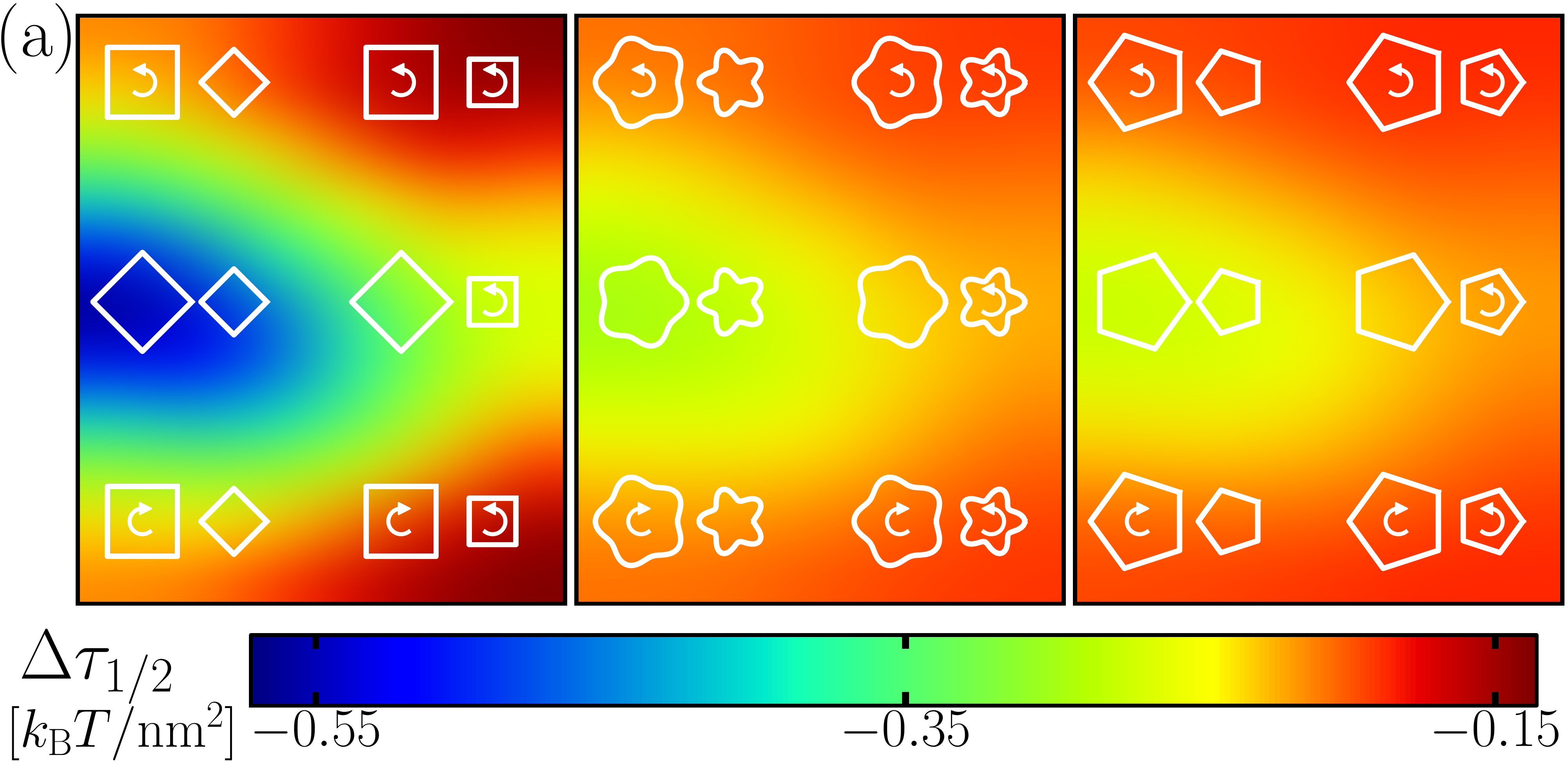}
 \onefigure[width=7.5cm]{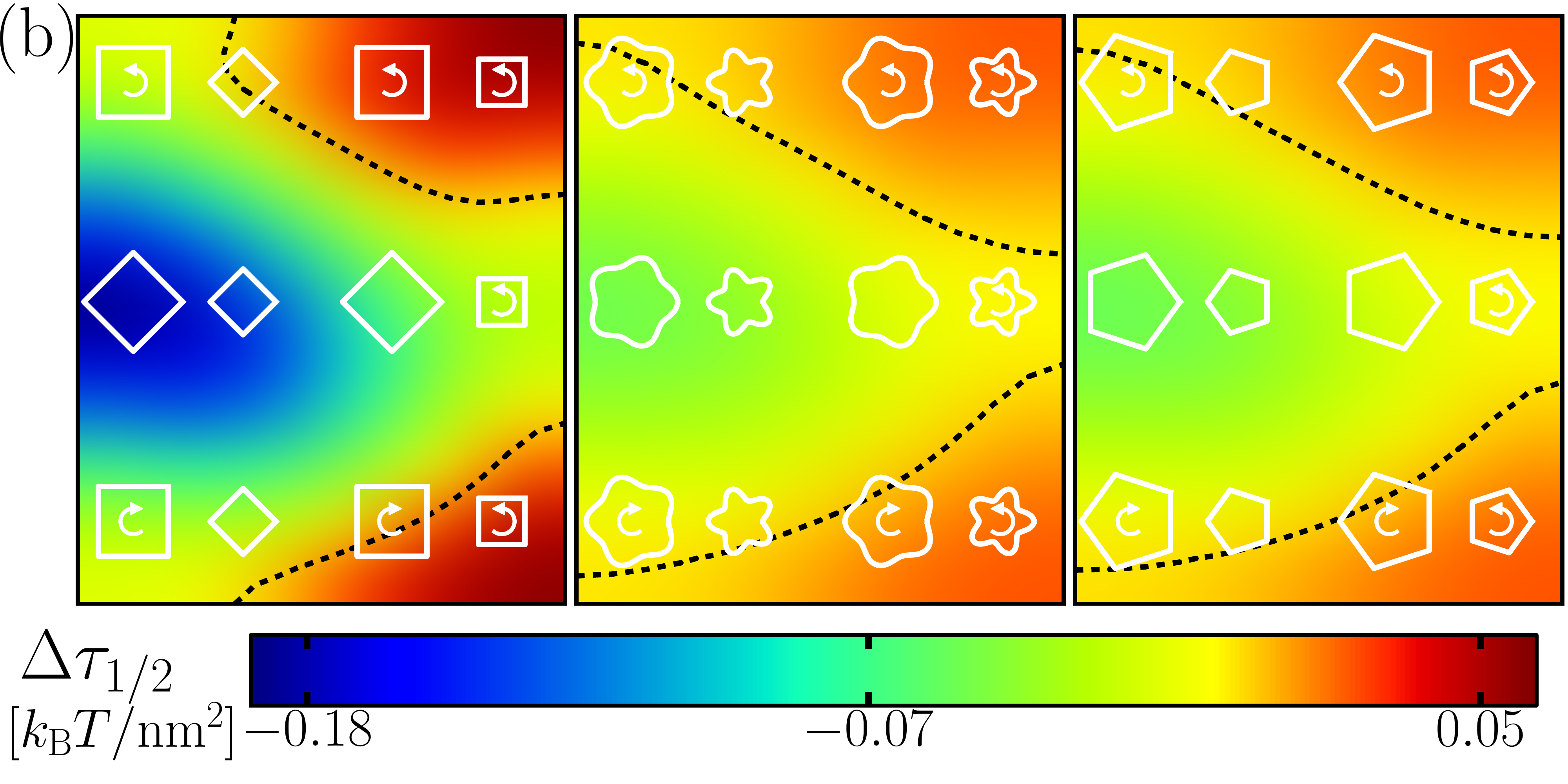}
\caption{(Color online)
Cooperative shift in gating tension, $\Delta \tau_{1/2}$, in eq.~(\ref{coopdef}) for tetrameric
MscL (left panels), the clover-leaf model of pentameric MscL (middle panels),
and the polygonal model of pentameric MscL (right panels) as a function of 
rotations of the right (horizontal axes) and left (vertical axes) channels
at (a) $d=9$~nm and (b) $d=10$~nm. 
Curved arrows show the directions of channel rotation relative to the tip-on reference configuration (midpoints on left axes).
Dashed lines indicate vanishing shifts in gating tension.
\label{fig:gating1}}  
\end{figure}

\begin{figure}[t]
\onefigure[width=7.5cm]{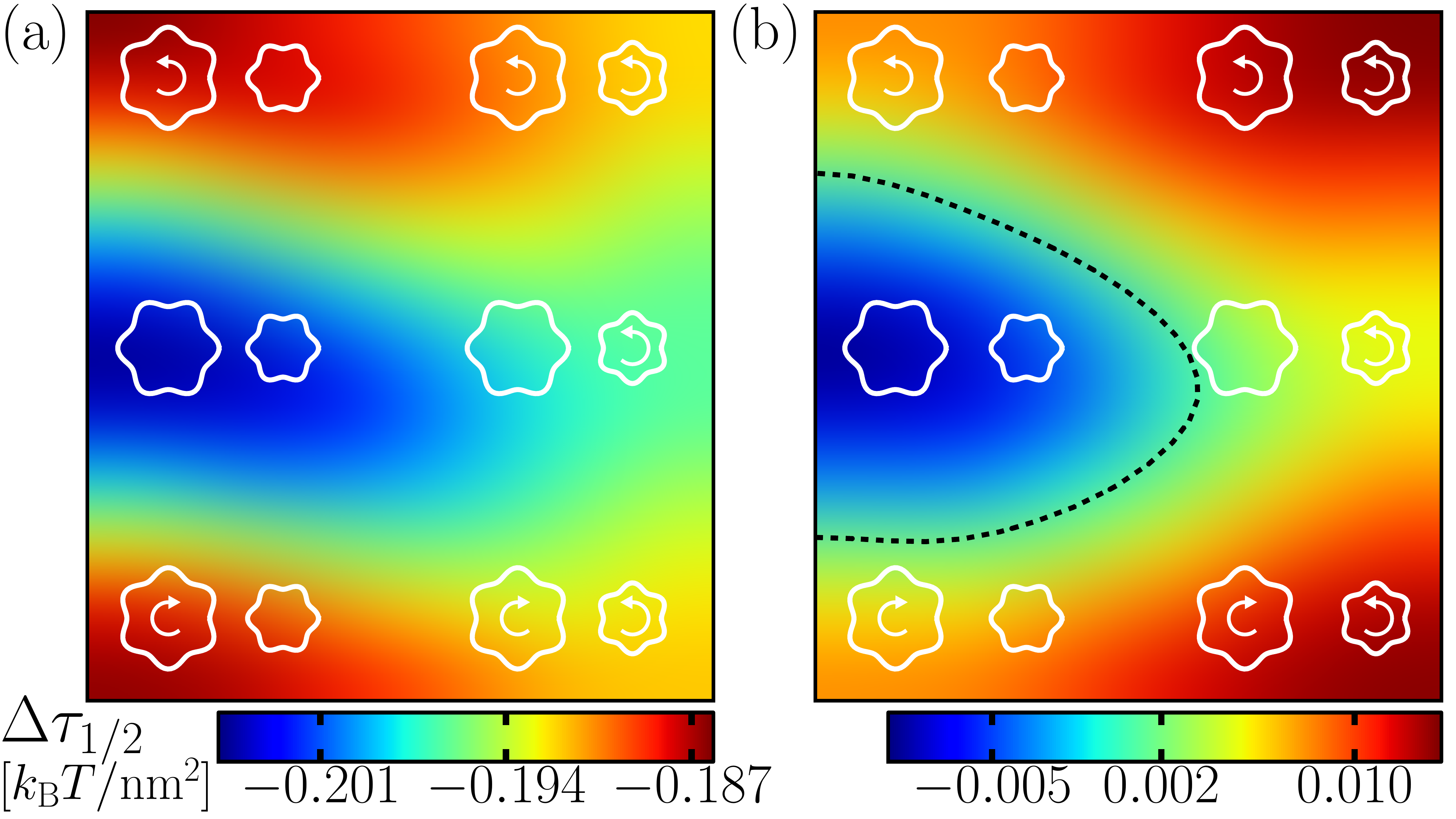}
\caption{(Color online)
Cooperative shift in gating tension, $\Delta \tau_{1/2}$, in eq.~(\ref{coopdef})
for MscL hexamers as a function of protein orientation at (a) $d=9$~nm and (b) $d=10$ nm. We use labeling conventions analogous to fig.~\ref{fig:gating1}. 
 \label{fig:hh}}
\end{figure}

\section{Anisotropic cooperativity} 
Membrane-mediated interactions can yield cooperative gating of mechanosensitive
ion channels \cite{Ursell2007,grage2011bilayer,Nomura2012,CAH2013a}. Since, as discussed above, the anisotropy of thickness deformations bears a signature of MscL structure, membrane-mediated cooperativity is expected to depend on the hydrophobic shape of MscL. We quantify membrane-mediated cooperativity
by calculating the opening probability in eq.~(\ref{eq:proba}) for a channel in close proximity to an already open channel. Membrane-mediated interactions
will shift the gating tension by
\begin{align} \label{coopdef}
\Delta \tau_{1/2}(\omega_1, \omega_2;d) = \tau_{1/2} (\omega_1, \omega_2;d) - \tau_{1/2}(d \to \infty)
\end{align}
for each proposed structure of MscL \cite{Liu2009,Chang1998,Sukharev2001a,Sukharev2001b,Saint1998},
where $\tau_{1/2}(d \to \infty)$ corresponds to the gating
tension of non-interacting MscL plotted in fig.~\ref{fig:gating_one}.

Figure~\ref{fig:gating1} shows $\Delta \tau_{1/2}$ for tetrameric MscL and the clover-leaf and polygonal models of pentameric MscL at two different channel separations, $d=9$ nm (fig.~\ref{fig:gating1}(a)) and $d=10$ nm (fig.~\ref{fig:gating1}(b)), for all relative orientations of MscL. The horizontal axes of the contour plots correspond to rotations of the right channels as indicated by arrows, while the vertical axes correspond to rotations of the left channels,
using the tip-on configuration as the reference state $(\omega_1, \omega_2) = (0,0)$.  For two open tetramers at $d=9$ nm, membrane-mediated interactions are highly favorable for the tip-on configuration, with an interaction strength $\sim\!\!10 \, k_\text{B}T$. Hence, we find strong cooperative effects for this configuration, lowering the gating tension by $\sim\!\!0.6 \,k_\text{B}T/$nm$^2$. Any deviation from the tip-on configuration weakens the cooperative effects: for instance, for the face-on configuration, the cooperative shift in gating tension only amounts to $\Delta \tau_{1/2} \approx  -0.15 \,k_\text{B}T/$nm$^2$. 
Similarly, we find that for both clover-leaf (fig.~\ref{fig:gating1}(a) middle panel) and polygonal (fig.~\ref{fig:gating1}(a) right panel) models
of pentameric MscL the tip-on configuration produces the greatest strength of cooperative interactions at small separations, yielding a decrease in gating tension by up to $\sim\!\!0.3 \,k_\text{B}T/$nm$^2$.

For increased channel separations, $d=10$ nm (fig.~\ref{fig:gating1}(b)), cooperative interactions between MscL are qualitatively different and weaker
in magnitude. In particular, for configurations close to the face-on configuration in parameter space (bottom and upper right corners of contour plots) membrane-mediated interactions now yield an increase in gating tension. Outside these regions, membrane-mediated interactions lead to a decrease in gating tension, by up to $\Delta \tau_{1/2}\approx-0.18 \,k_\text{B}T/$nm$^2$ for tetrameric MscL. As for small
channel separations, the strength of cooperative interactions is greater for tetrameric MscL (fig.~\ref{fig:gating1}(b) left panel) than for pentameric clover-leaf and polygonal models of MscL
(fig.~\ref{fig:gating1}(b) middle and right panels).

The hexameric model of MscL yields cooperative effects which, in some respects, deviate from the trends noted above (fig.~\ref{fig:hh}). At small separations, the least cooperative configuration is not the face-on orientation as found for tetramers and pentamers (fig.~\ref{fig:gating1}), but the face-tip
configuration (lower and upper right corners in fig.~\ref{fig:hh}(a)).
As the channel separation is being increased (fig.~\ref{fig:hh}(b)), this configuration yields a gating tension which is increased compared to non-interacting MscL but, similarly to tetrameric and pentameric MscL, the maximum increase in gating tension occurs for the face-on configuration of MscL hexamers. Since MscL hexamers have a higher-order symmetry than tetrameric and pentameric MscL, the predicted anisotropy in the cooperative gating of MscL is least pronounced for MscL hexamers (and most pronounced for MscL tetramers).

\begin{figure}[t]
\onefigure[width=7.5cm]{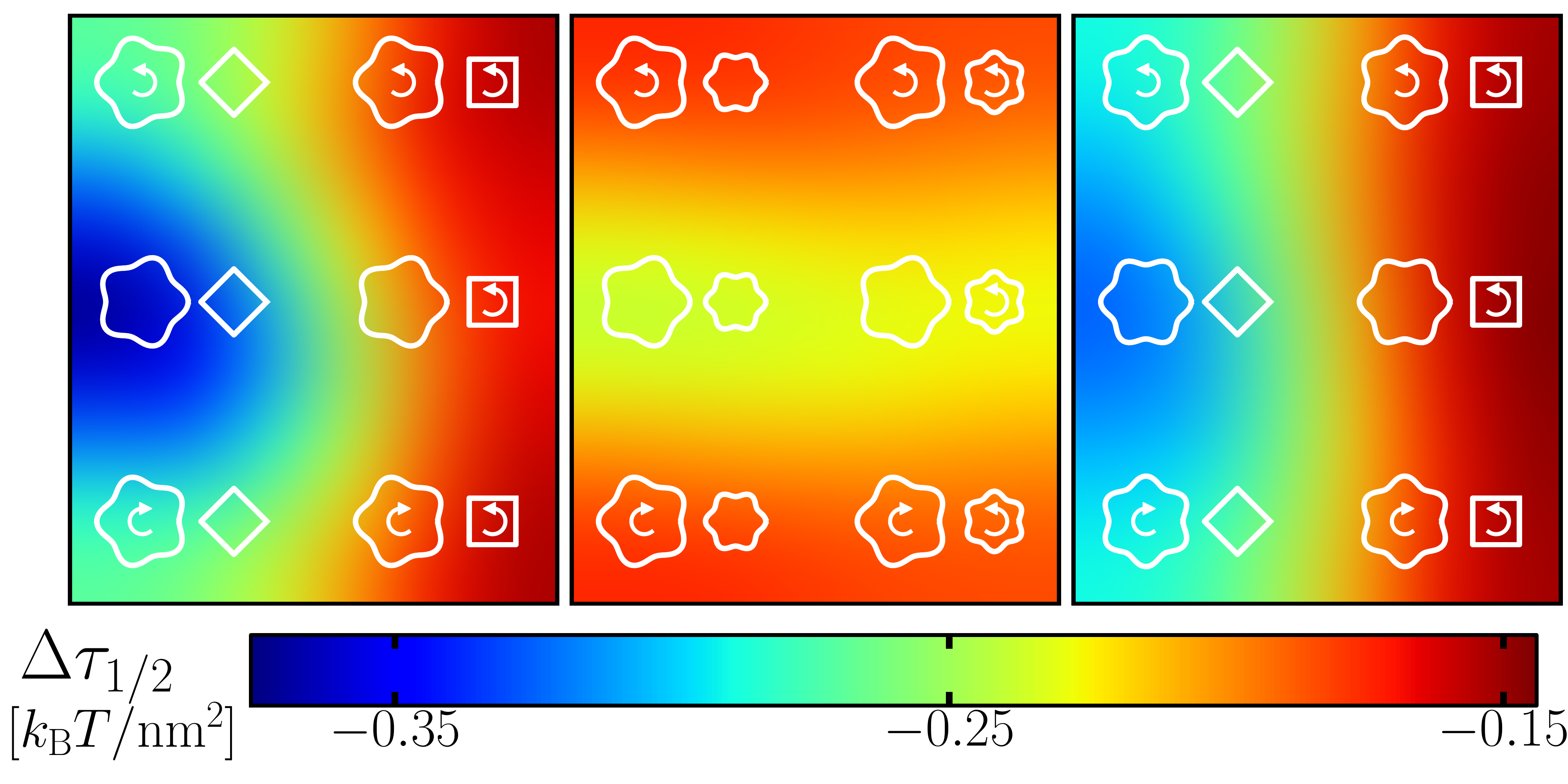}
\caption{(Color online)
Cooperative shift in gating tension, $\Delta \tau_{1/2}$, in eq.~(\ref{coopdef})
at $d=9$~nm for interactions between distinct oligomeric states of MscL. The left and middle panels correspond to the gating of a closed MscL tetramer and a closed MscL hexamer in proximity to an open clover-leaf MscL pentamer. The right panel corresponds to the gating of a closed MscL tetramer in proximity
to an open MscL hexamer. We use labeling conventions analogous to fig.~\ref{fig:gating1}. 
 \label{fig:mixte}}  
\end{figure}

Finally, in fig.~\ref{fig:mixte} we consider cooperative effects between
distinct oligomeric states of MscL. For closed MscL tetramers in close proximity
to open clover-leaf pentamers (fig.~\ref{fig:mixte} left panel) the shift in gating tension varies between $-0.4$~$k_\text{B}T/$nm$^2\lessapprox\Delta \tau_{1/2}\lessapprox-0.2$~$k_\text{B}T/$nm$^2$. We find that the variation in $\Delta \tau_{1/2}$ is most pronounced along the horizontal axis, which
can be understood by comparing the color distributions in the left and middle panels of fig.~\ref{fig:gating1}: rotation of a tetramer has a more pronounced effect on cooperative interactions than rotation of a pentamer, thus creating the preferred rotation direction in the left panel of fig.~\ref{fig:mixte}. Similarly, the orientational anisotropy in the cooperative gating of a MscL
hexamer due to an open clover-leaf pentamer (fig.~\ref{fig:mixte} middle panel)
is most pronounced for rotations of the pentamer, yielding a horizontal band
in the middle panel of fig.~\ref{fig:mixte}. Furthermore, the anisotropy in
the cooperative gating of a MscL pentamer or a MscL tetramer (fig.~\ref{fig:mixte} right panel) in close proximity to an open MscL hexamer is dominated
by rotations of the channel of lower-order symmetry. This behavior becomes
more pronounced with decreasing symmetry of the lower-order channel and,
hence, increasing deviations from the perfectly isotropic cylinder model
of MscL.

\section{Conclusion}
In this letter we carried out a systematic
survey of membrane-mediated interactions and cooperative gating for a variety
of structural models proposed for MscL \cite{Liu2009,Chang1998,Sukharev2001a,Sukharev2001b,Saint1998}.
We predict that the molecular structure of MscL is reflected in the structure of membrane-mediated interactions, yielding characteristic sequences of gateway states for the dimerization of MscL. Furthermore, we find substantial shifts in the cooperative gating tension of MscL with the oligomeric state of MscL, and predict general relations between MscL stoichiometry, spatial arrangement, and cooperativity. Thus, our results suggest that the spatial arrangement and function of MscL bear distinctive signatures of MscL structure, which may provide novel approaches for the experimental
dissection of the MscL oligomeric states and shapes most relevant \textit{in
vivo} \cite{Haswell2011,Dorwart2010,Iscla2011,Gandhi2010}. The modeling approach developed here establishes a quantitative relation between the observed shapes and cooperative function of membrane proteins.

\acknowledgments
This work was supported at USC by the National Science Foundation through NSF award number DMR-1206332 and by the USC Center for High-Performance Computing and Communications, and at UCLA by the National Science Foundation through NSF award numbers CMMI-0748034 and DMR-1309423. CAH acknowledges additional support through the Aspen Center for Physics and the National Science Foundation under award number PHY-1066293.


\end{document}